\pdfoutput=1
\documentclass[a4paper,american,aps,citeautoscript,floatfix,pdftex,pra,superscriptaddress,twocolumn]{revtex4-1}

\usepackage{amsfonts,amsmath,amssymb}
\usepackage{graphicx}
\usepackage{hypernat}
\usepackage[T1]{fontenc}
\usepackage[utf8]{inputenc}
\usepackage{mathptmx}
\usepackage{soul}
\usepackage{subfigure}
\usepackage{textcomp}
\usepackage[disable,textsize=tiny,textwidth=9mm]{todonotes}
\usepackage{xspace}
\usepackage{hyperref}

\newlength{\figwidth}
\newcommand{\cost}{\ensuremath{\langle\cos^2\theta\rangle_\text{2D}}\xspace}
\newcommand{\costhreeD}{\ensuremath{\langle\cos^2\theta\rangle}\xspace}
\newcommand{\degree}[1]{\xspace\ensuremath{^\circ}#1\xspace}%
\newcommand{\ie}{i.\,e.}%

\newcommand{\Iplus}{\ensuremath{\text{I}^+}\xspace}%

\newcommand{\Nuptot}{\ensuremath{\text{N}_{\text{up}}/\text{N}_{\text{tot}}}\xspace}

\newcommand{\cfeldesy}{\affiliation{Center for Free-Electron Laser Science, DESY, Notkestrasse 85, 22607 Hamburg, Germany}}%
\newcommand{\uhhcui}{\affiliation{The Hamburg Centre for Ultrafast Imaging, Luruper Chaussee 149, 22761 Hamburg, Germany}}%
\newcommand{\uhhphys}{\affiliation{Department of Physics, University of Hamburg, Luruper Chaussee 149, 22761 Hamburg, Germany}}%

\begin{document}
\title{Strongly aligned and oriented molecular samples at a kHz repetition rate}
\author{Sebastian Trippel}%
\author{Terry Mullins}%
\author{Nele L.\,M.\ Müller}%
\author{Jens S.\ Kienitz}%
\author{Karol Długołęcki}%
\cfeldesy
\author{\mbox{Jochen Küpper}}%
\email[]{jochen.kuepper@cfel.de}%
\homepage{http://desy.cfel.de/cid/cmi}%
\cfeldesy\uhhphys\uhhcui%
\date{\today}%
\begin{abstract}\noindent%
   \begin{center}
      \vspace*{-6mm}
      \emph{Dedicated to Bretislav Friedrich on the occasion of his 60$^\text{th}$ birthday}
   \end{center}
   \noindent We demonstrate strong adiabatic laser alignment and mixed-field orientation at kHz
   repetition rates. We observe degrees of alignment as large as $\cost=0.94$ at 1~kHz operation for
   iodobenzene. The experimental setup consist of a kHz laser system simultaneously producing pulses
   of 30\,fs (1.3\,mJ) and 450\,ps (9\,mJ). A cold 1\,K state-selected molecular beam is produced at
   the same rate by appropriate operation of an Even-Lavie valve. Quantum state selection has been
   obtained using an electrostatic deflector. A camera and data acquisition system records and
   analyzes the images on a single-shot basis. The system is capable of producing, controlling
   (translation and rotation) and analyzing cold molecular beams at kHz repetition rates and is,
   therefore, ideally suited for the recording of ultrafast dynamics in so-called ``molecular
   movies''.
\end{abstract}
\maketitle

\section{Introduction}
\label{sec:introduction}

Aligned and oriented molecules serve as ideal samples to study steric effects in chemical
reactions~\cite{Brooks1976, Stapelfeldt:RMP75:543} and to image the structure and dynamics of
complex molecules directly in the molecular frame, if that is strongly confined, \ie, linked to the
laboratory frame of the measurement. This would yield so-called ``molecular movies'' of the ongoing
dynamics, conceivably without prior knowledge on the investigated system.

Bretislav Friedrich has been at the forefront of the development of methods to control complex
molecules, including brute force orientation~\cite{Friedrich:Nature353:412, Loesch:JCP93:4779},
laser alignment~\cite{Friedrich:PRL74:4623} and mixed-field orientation~\cite{Friedrich:JCP111:6157,
   Friedrich:JPCA103:10280}. At that time, the degree of alignment and orientation was too weak to
image molecular dynamics directly in the molecular frame. However, over the last two decades, the
available degree of control has been constantly increased. The combination with rotational-state
selection~\cite{Holmegaard:PRL102:023001, Putzke:PCCP13:18962} has improved the achievable control
dramatically, with the strongest demonstrated degree of alignment so far of
$\cost>0.97$~\cite{Holmegaard:PRL102:023001}. In addition, recent results on the orientation of
molecules in mixed fields show that the adiabaticity and the resulting degree of orientation
strongly depend on the applied electric fields, \ie, the laser-pulse
duration~\cite{Nielsen:PRL108:193001}.

Strong laser alignment and mixed-field orientation has been exploited in the recording of molecular
frame photoelectron angular distributions of complex molecules~\cite{Holmegaard:NatPhys6:428}.
Controlled samples increase the contrast in all direct imaging experiments. They allow the simple
averaging of many individual experiments, they simplify the data analysis, and no orientation
relationship between patterns from randomly oriented molecules need to be derived
numerically~\cite{Spence:PRL92:198102, Filsinger:PCCP13:2076}. Moreover, they are crucial to various
advanced ``photography'' experiments: Tomographic reconstruction approaches for
X-ray~\cite{Pabst:PRA81:043425, Filsinger:PCCP13:2076} or electron
diffraction~\cite{Reckenthaeler:PRL102:213001, Hensley:PRL109:133202} and photoelectron holography
experiments of aligned molecules require typically $\cost\approx0.9$~\cite{Filsinger:PCCP13:2076,
   Krasniqi:PRA81:033411}.

Such a strong degree of alignment has, so far, only been achieved in adiabatic alignment experiments
making use of very cold molecular beams and nanosecond laser pulses at a repetition rate of a few
10~Hz~\cite{Kumarappan:JCP125:194309, Holmegaard:PRL102:023001}. However, the low repetition rates
renders time resolved studies of molecular dynamics, with their generally low count rates, tedious,
at least, or even infeasible. Therefore, experimental setups which provide strong alignment at high
repetition rates are highly desirable. This requires the production of cold molecular beams of
complex molecules and strong laser fields with pulse durations that are comparable or longer than
the rotational period of the molecules. The lack of nanosecond lasers with sufficient pulse energies
at kHz repetition rates suggests the use of laser pulses generated by an amplified Ti:Sa laser
system. However, the generation of such laser pulses with pulse durations on the order of 1~ns and
the required peak intensities of $>10^{11}$~W/cm$^3$ is challenging. Moreover, \emph{a priori} it
has been unclear whether ``heavy'' molecules with a rotation time in the order of a few hundred
picoseconds (like iodobenzene) can be strongly aligned by a sub-nanosecond, strongly linearly
chirped broadband 800\,nm laser pulse.

Recently, some relevant molecular beam setups with high repetition rates have been developed and
some individual ingredients of the necessary control and detection details have been
demonstrated~\cite{Horio:RSI80:013706, Ren:PRA85:033405, Irimia:RSI80:3958}. A benchmark experiment
demonstrating long-pulse alignment of molecules in a low-pressure continuous beam at 1~kHz
repetition rate, probed with short ps x-ray pulses from a synchrotron source, has demonstrated weak
alignment of $\costhreeD\approx0.4$~\cite{Peterson:APL92:094106}. Impulsive alignment experiments
have been performed, again exploiting continuous molecular beams, with lasers operating at kHz
repetition rates~\cite{Peronne:PRL91:043003, Rouzee:JPB45:074016}. The achieved degree of alignment
however is typically also only moderately strong ($\costhreeD\le0.85$). This makes these approaches
not very well suited for molecular-frame imaging studies of complex molecules. Moreover, in these
experiments the alignment typically only persists for short periods of time ($\sim\!1$~ps), which
severely limits the time-window for time-resolved experiments, esp.\ for large amplitude dynamics,
such as conformer interconversion or folding motions.

Here, we present a new experimental setup that provides strongly aligned and oriented samples of
state-selected molecules at a repetition rate of 1~kHz. This rate is a good compromise between
current table-top laser systems, pulsed molecular beam sources, and high speed camera systems. In
addition, it demonstrates a clear pathway for the sample preparation at upcoming light sources with
high-repetition rates, such as x-ray free-electron lasers (XFELs), synchrotrons, and laser based
high-harmonic generation (HHG) sources. In order to efficiently use these light sources, the
capability of high repetition rate adiabatic alignment and orientation is highly desirable. Whereas
the availability of synchronized pulses from high-power table-top laser systems is practically an
intrinsic feature, XFEL facilities, such as the European XFEL in Hamburg, are actively pursuing the
setup of high-repetition rate lasers that meet the requirements set by the current work.

\section{Experimental setup}
\label{sec:setup}

\begin{figure}
   \centering
   \includegraphics[width=\linewidth]{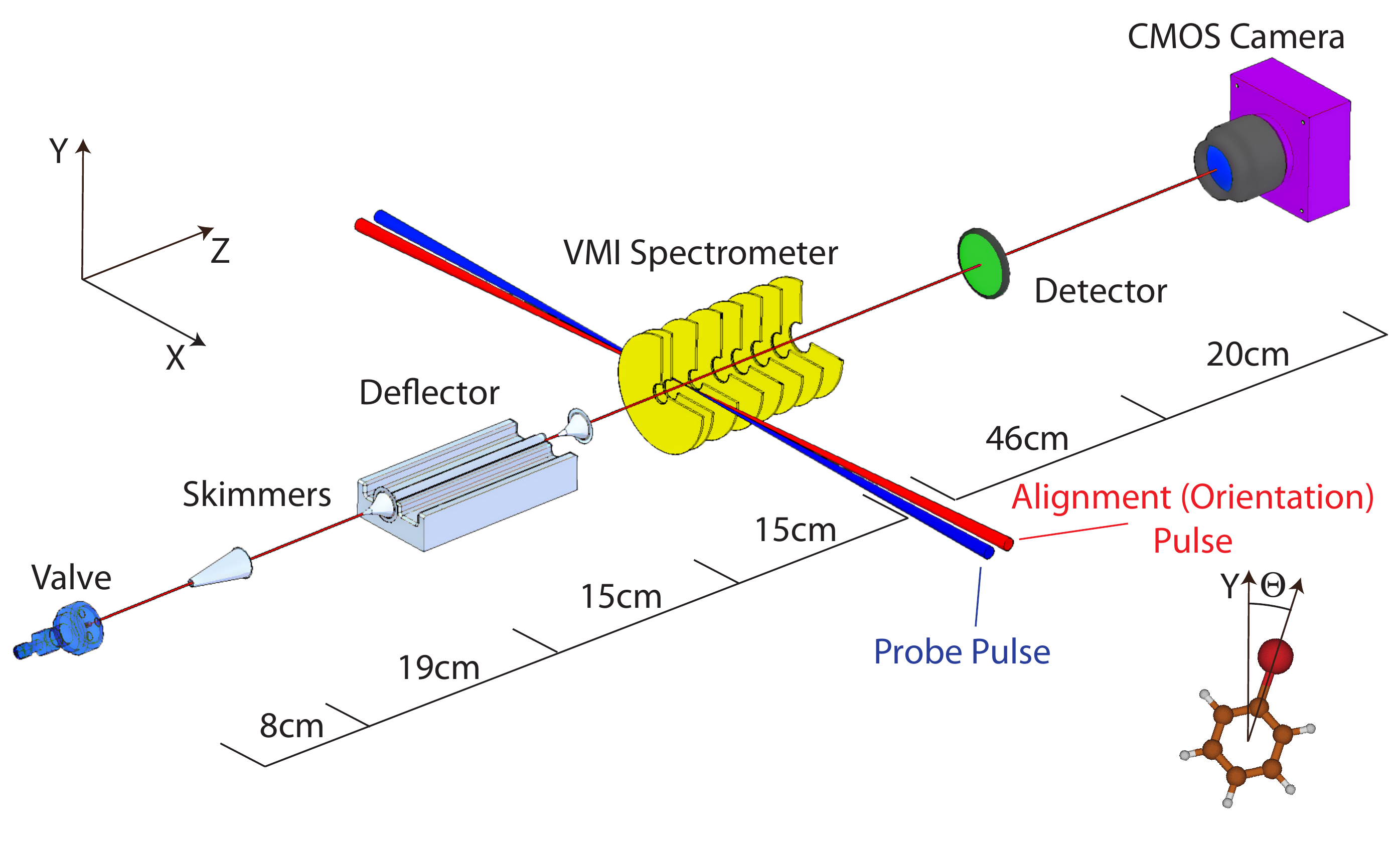}
   \caption{Schematic of the experimental setup. The pulsed molecular beam passes two skimmers
      before it enters the electrostatic deflector where it is dispersed, depending on its quantum
      states, along the Y-axis. The alignment or orientation laser pulses as well as the probe pulse
      cross the molecular beam inside of a velocity map imaging spectrometer. The ions are mapped on
      a position sensitive detector consisting of an MCP and a phosphor screen. The angle
         $\theta$ is defined as the angle between the Y-axis and the most-polarizable axis of the
         molecule, \ie, the C--I bond axis.}
   \label{fig:setup}
\end{figure}

The schematic of the experimental setup is shown in \autoref{fig:setup}.
Details will be published in a longer account and only a brief description will be presented here. A
pulsed molecular beam is provided by expanding 10~mbar of iodobenzene seeded in 120~bar of helium
through an Even-Lavie valve~\cite{Even:JCP112:8068} cooled to -20\degree{C}. After passing two
skimmers the molecular beam enters an electric deflector, where the molecules are dispersed
according to their quantum state~\cite{Filsinger:JCP131:064309}. The state selected molecular
ensembles are aligned or oriented by laser or mixed dc-electric and laser
fields~\cite{Friedrich:JCP111:6157, Friedrich:JPCA103:10280, Holmegaard:PRL102:023001,
   Ghafur:NatPhys5:289}, respectively, inside a velocity map imaging spectrometer
(VMI)~\cite{Eppink:RSI68:3477}. The angular confinement is probed through strong-field multiple
ionization by a short laser pulse followed by Coulomb explosion of the molecule. The resulting I$^+$
ions are velocity mapped onto a 40\,mm diameter position sensitive detector (Photonis) consisting of
a multi-channel-plate (MCP), a fast phosphor screen (P-46), and a high frame-rate camera. High speed
oil-free pumping (4000~l/s for the source and 2000~l/s for deflection and detection chambers,
respectively) and optimized operation conditions of the Even-Lavie pulsed valve allow for the
generation of dense ($>10^{9}$~molecules/cm$^3$ in the interaction volume) and cold (1~K) molecular
beams, whose population distribution is further reduced to some ten rotational states by
state-selection~\cite{Filsinger:JCP131:064309}.

Alignment and ionization laser pulses are provided by an amplified femtosecond laser system
(Coherent Legend Elite Duo HE USX NSI) at a repetition rate of 1~kHz. The total output power of
the system is larger than 10~W with a bandwidth of $\ge72$~nm centered at 800~nm. Directly behind
the amplification stages the laser beam is split into two parts, an alignment (orientation) beam
($\approx9$~mJ/pulse) and a probe beam ($\approx1.3$~mJ/pulse). The duration of the alignment pulse
can be compressed or stretched (negatively chirped) with an external compressor continuously from
40\,fs up to 520~ps. The probe beam is compressed to 30~fs using the standard grating based
compression setup. Since both beams are produced by the same amplifier system they are inherently
synchronized. The two beams are incident on a 60~cm focal length lens parallel to each other with a
transverse distance of 10~mm, resulting in field strength of up to $5\times10^{11}$~W/cm$^2$ and
$5\times10^{14}$~W/cm$^2$ for alignment and probe pulse, respectively. The foci are overlapped in
space and time in the molecular beam and in the center of the velocity map imaging spectrometer.
Vertically scanning the lens allows probing different parts of the, typically, quantum state
dispersed, molecular beam, \ie, to probe ensembles with varying rotational excitation and
correspondingly varying effective dipole moments and effective polarizabilities. This motion is
automatized and thus one can completely automatically measure the ion-distribution in the VMI as a
function of vertical beam position. From this data, one can determine the molecular beam density,
the degree of alignment, and the orientation all at once. A typical scan over the molecular beam at
1~kHz repetition rate takes about 1000~s (\emph{vide infra}).

The imaging system CMOS camera (Optronis CL600x2) is operated at a camera-link-readout-limited
resolution of $480\times480$ pixel at sustained 1~kHz repetition rate. This corresponds to a spatial
resolution of 80~\textmu{m} on the phosphor screen. The typical spatial illumination on the camera
corresponding to one ion is four pixels. The background corrected camera images are analyzed for
every single shot with a centroiding algorithm on a standard PC computer, making use of eight available cores by sequentially distributing the images
onto different cores. The coordinates of the ion hits are passed to the main data acquisition
system. The single-shot analysis allows high signal rates exploiting the high saturation limit of
the detection system.

\section{Results}
\label{sec:results}

\subsection{1D Alignment}
\label{sec:1d-alignment}

\begin{figure}
   \centering%
   \includegraphics[width=\linewidth]{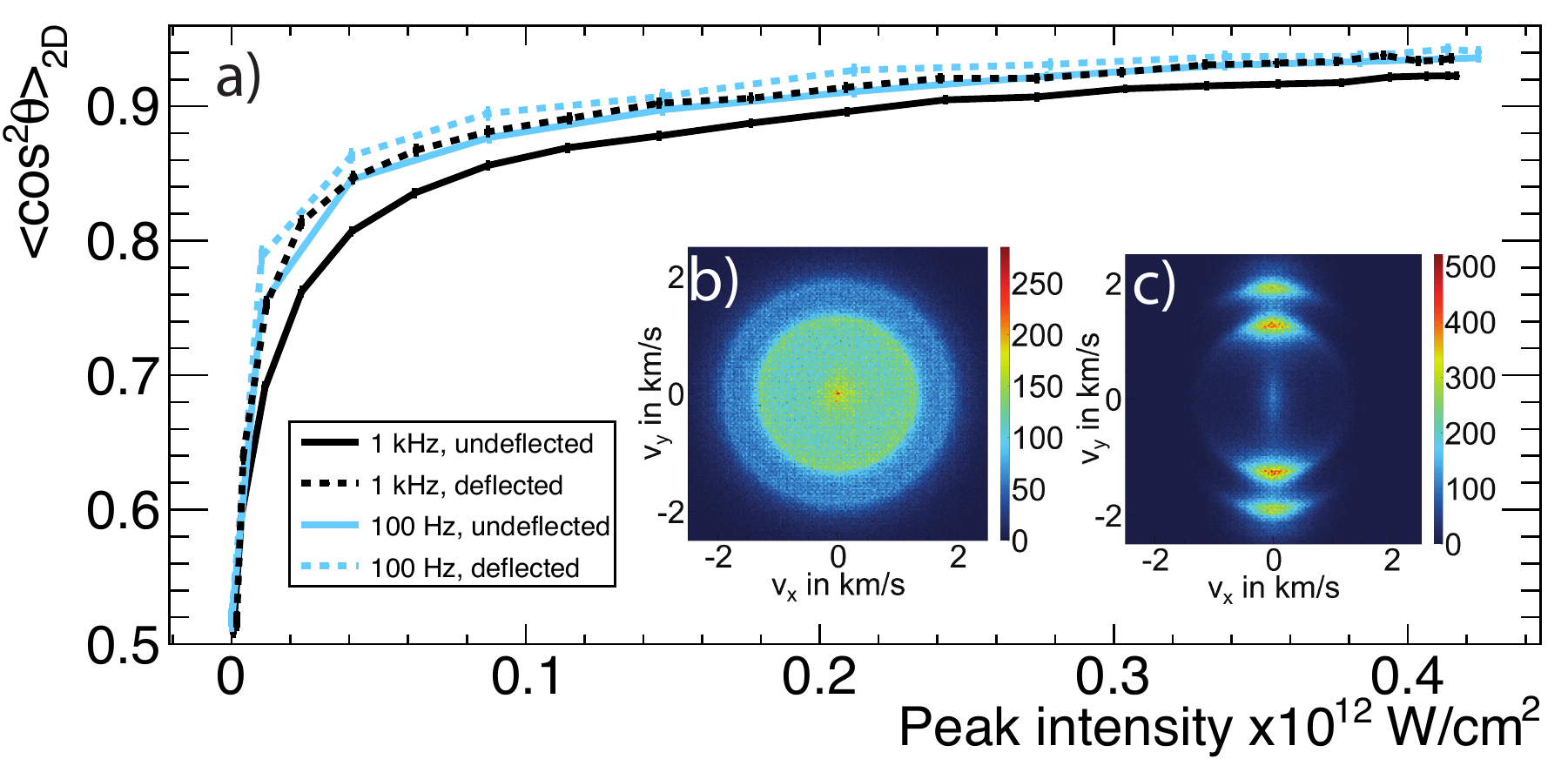}%
   \caption{(Color online). a) The degree of alignment as a function of the peak intensity of the
      alignment pulse for a repetition rate of 1~kHz in solid black (deflector off) and dashed black
      (deflected) as well as for a repetition rate of 100~Hz in solid blue/grey (deflector off) and
      dashed blue/grey (deflected). The insets show b) the 2D velocity distribution without
      alignment laser pulse and c) with an alignment pulse with a peak intensity of
      \mbox{$0.42\times10^{12}$~W/cm$^{2}$} at a repetition rate of 1~kHz in the deflected part of
      the beam.}
   \label{fig:1D-alignment}
\end{figure}

\autoref{fig:1D-alignment}\,a) shows the degree of alignment \cost~\footnote{The degree of
      alignment \cost is defined as $\int_0^{\pi}\int_{v1}^{v2} \cos^2\theta f(\theta,v) d\theta dv$
      with $f(\theta,v)$ being the two dimensional ion distribution on the detector. The angle
      $\theta$ is defined as the angle between the polarization of the alignment laser and the
      projection of the three dimensional ion velocity vector onto the two dimensional detector
      plane. The velocity is given by $v = \sqrt{v_x^2+v_y^2}$.} as a function of the peak intensity
   of the alignment pulse; see the inset of \autoref{fig:setup} for a definition of $\theta$.
The results are shown for a repetition rate of 1~kHz in solid black (deflector off) and dashed black
(deflected beam) as well as for a repetition rate of 100~Hz in solid blue/grey (deflector off) and
dashed blue/grey (deflected).
The observed power dependence of the degree of alignment for a cold beam is as
expected~\cite{Kumarappan:JCP125:194309}. For randomly aligned molecules a degree of alignment
   $\cost=0.5$ is expected. The degree of alignment increases with increasing laser intensity, but
   is limited by $\cost=1$. The 2D momentum image for I$^+$ ions for the alignment-field-free case
at 1~kHz repetition rate is shown in \autoref{fig:1D-alignment}\,b). The distribution is circularly
symmetric as expected for the case of an isotropic sample and the polarization of the probe laser
pulse linear and perpendicular to the detector plane. The degree of alignment is given by
$\cost=0.5007(5)$. \autoref{fig:1D-alignment}\,c) shows the corresponding ion distribution when the
molecules are aligned at 1~kHz repetition rate along the alignment pulse polarization axis, \ie,
linear, vertical and parallel to the detector plane. The number of ions is about three ions/pulse
and the image has been recorded in less than 10 minutes. The peak laser intensity is
   $4.2\times10^{11}$~W/cm$^2$. The degree of alignment determined from the outer structure (between
   $v_1$=1.7~km/s and $v_2$=2.2~km/s) of the two dimensional ion distribution is $\cost=0.935(1)$.
   The radial structures in the ion images indicate that there are two fragmentation channels
   present in the Coulomb explosion of iodobenzene: the inner ring corresponds to \Iplus
   recoiling from a singly charged phenyl, whereas the outer ring corresponds to \Iplus
   recoiling from doubly charged phenyl~\cite{Larsen:JCP111:7774}. Using the outer ring for
   the determination of \cost is favorable as it corresponds to the most sudden fragmentation
   channel and, therefore, to the best axial recoil conditions. The derived value underestimates the
   degree of alignment since the probe and alignment pulse have perpendicular polarization and,
   therefore, the least-aligned molecules are ionized with the highest
   efficiency~\cite{Kumarappan:JCP125:194309}.

A slightly higher degree of alignment is obtained when the valve is operated at 100~Hz repetition
rate. The degree of alignment with the deflector off at 100\,Hz repetition rate is comparable to the
one obtained at 1\,kHz repetition rate in the deflected beam. A slight increase of the degree of
alignment to $\cost=0.942(1)$ is observed in the deflected part of the beam at 100\,Hz repetition
rate. In order to investigate the reason for this behavior we have operated the valve at 100\,Hz at
a higher temperature and leaked helium gas into the chamber. Both, temperature and source chamber
pressure, have been adjusted to match the conditions found at 1~kHz repetition rate. Under these
conditions we observed a degree of alignment that matches the 1~kHz results. Therefore, the slightly
smaller degree of alignment is attributed to a higher valve temperature and a higher background
pressure in the source chamber when the valve is operated at 1~kHz. The lower peak intensity of the
employed alignment pulse is the reasons for the smaller value of $\cost$ compared to previous
reported best value obtained using a 20~Hz 10~ns injection seeded Nd:YAG
laser~\cite{Holmegaard:PRL102:023001}. Obviously, these experimental limits could be solved by
exploiting higher-power lasers and considerably larger vacuum pumps, which have not been available
for this study.

   The pulse length of the alignment pulse is 450~ps. The rotational period of iodobenzene (J-type
   revival) is 707.7~ps~\cite{Poulsen:JCP121:783, Dorosh:JMolSpec246:228}. Thus, the laser
   pulse length is shorter than the rotational period of iodobenzene, placing this study in the
   intermediate regime between adiabatic and impulsive alignment -- closer to the adiabatic case.
   The observed degree of alignment clearly indicates comparable control as in previously reported
   adiabatic alignment experiments~\cite{Larsen:JCP111:7774, Holmegaard:PRL102:023001}. The
   detailed influence of non-adiabatic effects for the studied system is not clear and a more
   detailed investigation of the alignment dynamics as a function of the laser pulse duration is in
   preparation.

\subsection{1D Orientation}
\label{sec:1d-orientation}
\begin{figure}
   \centering%
   \includegraphics[width=\linewidth]{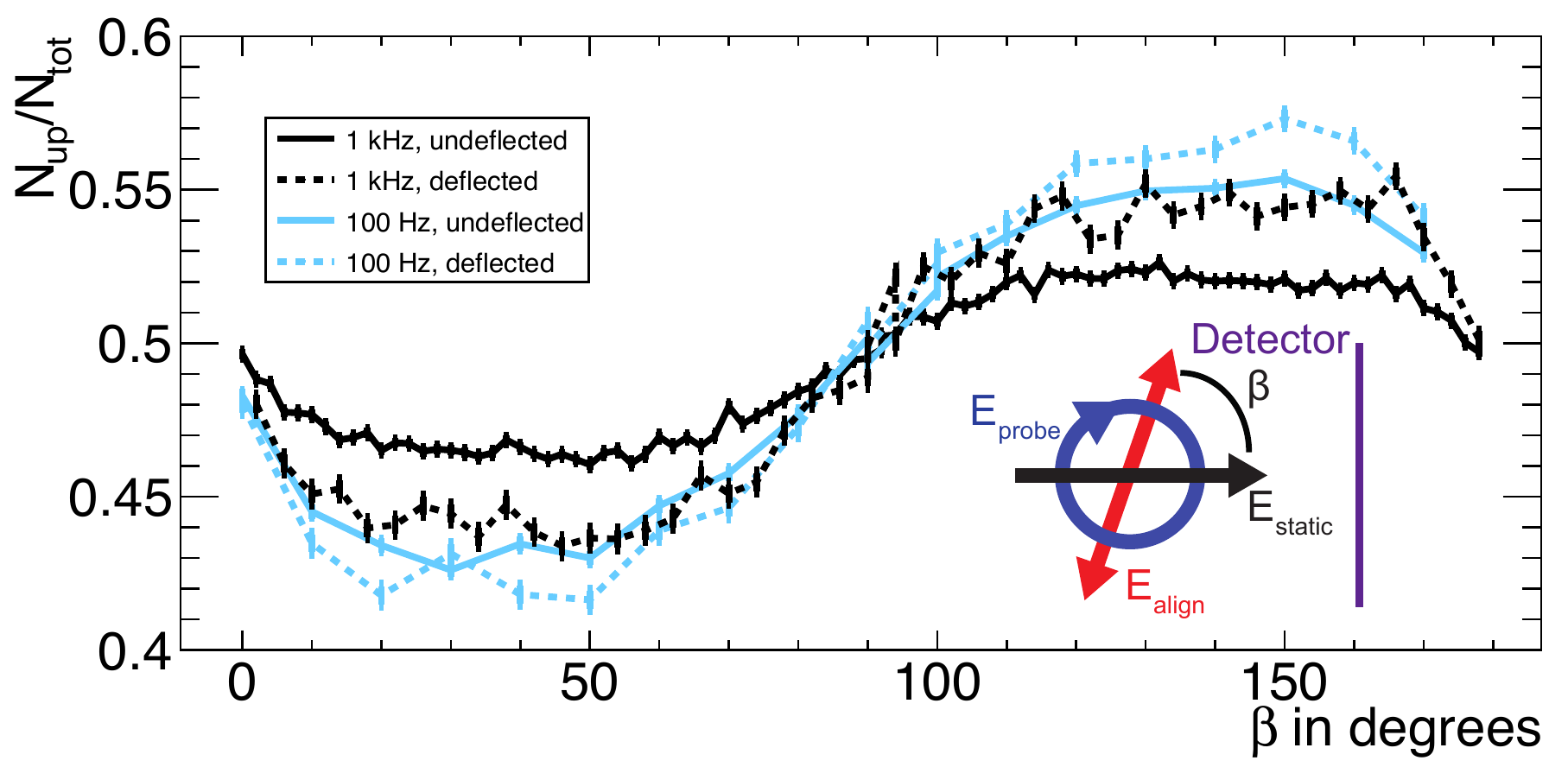}%
   \caption{(Color online) The ratio \Nuptot as a function of the angle $\beta$. The color scheme is
      as in \autoref{fig:1D-alignment}\,a). The inset shows the definition of the angle $\beta$. The
      typical time scale of one full $\beta$-angle scan at 1~kHz is in the order of 1000~s.}
   \label{fig:1D-orientation}
\end{figure}

\autoref{fig:1D-orientation} shows the ratio of ions in the upper half of the detector images
divided by the total number of ions \Nuptot\ as a function of the angle $\beta$.
The color scheme is as in \autoref{fig:1D-alignment}a). The angle $\beta$ is defined as the angle
between the polarization of the alignment laser and the static electric field of the VMI
spectrometer, $\text{E}_\text{static}=840$~V/cm, as shown in the inset
of \autoref{fig:1D-orientation}. The probe beam is circularly polarized. The observed dependence of
the orientation as a function of the angle beta is as expected from previous low-repetition rate
experiments~\cite{Holmegaard:PRL102:023001, Filsinger:JCP131:064309}. The molecules are better
oriented in the deflected part of the beam, where the population is confined to the energetically
lowest, \ie, the most polar, rotational states. In addition we observe better orientation at 100~Hz
than at 1~kHz repetition rates. The reason for this is the slightly increased rotational temperature
of the 1~kHz molecular beam, as discussed above. In comparison with the orientation for the same
molecule obtained using 10~ns alignment-laser pulses~\cite{Holmegaard:PRL102:023001} we obtain a
considerably smaller degree of orientation in the current experiment. This can be attributed to the
nonadiabatic mixing of levels in the near-degenerate doubles created by the strong laser field,
which has been shown to be more prominent for shorter laser pulses~\cite{Nielsen:PRL108:193001}.
This is also in agreement with earlier theoretical studies that have shown that the degree of
impulsive orientation, using short laser pulses, is limited by the magnitude of the applied DC
electric field~\cite{Rouzee:NJP11:105040}.

\subsection{Molecular-beam deflection dependence}
\label{sec:lens-dependence}

\begin{figure}
   \centering%
   \includegraphics[width=\linewidth]{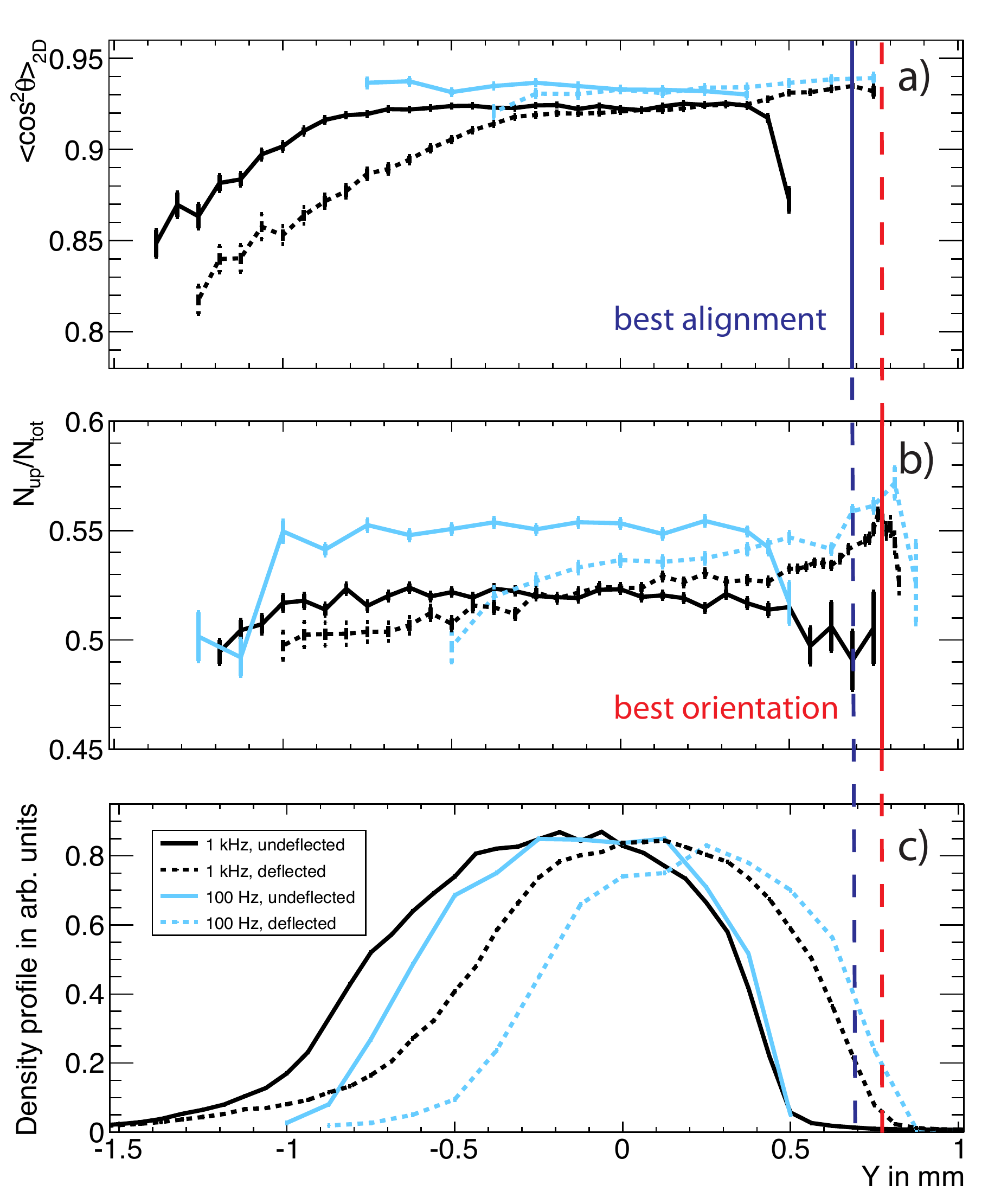}%
   \caption{(Color online) The degree of alignment (a), the ratio \Nuptot (b) and the molecular
         beam density (c) as a function of the position in the molecular beam. The color scheme is
      as in \autoref{fig:1D-alignment}a). The positions for the best alignment and orientation are
      marked by vertical lines. The time for the data acquisition at 1\,kHz for a single curve is in
      the order of half an hour.}
   \label{fig:lens-dependence}
\end{figure}

 In \autoref{fig:lens-dependence} the vertical molecular beam profiles
   (\autoref{fig:lens-dependence}\,c) and the dependence of the degrees of alignment
   (\autoref{fig:lens-dependence}\,a) and orientation (\autoref{fig:lens-dependence}\,b)
   on the position in the molecular beam, measured in a single lens scan as described above, are
shown.
The color scheme of the plots is the same as in \autoref{fig:1D-alignment}a). The molecular beam
profiles and the observed deflection, for an applied deflector voltage of 12~kV, are in good
agreement with previous measurements~\cite{Filsinger:JCP131:064309}. For the beam with the deflector
off, the observed degrees of alignment and orientation are practically constant over the main part
of molecular beam profile. Towards the sides, however, the achievable control decreases, which is
especially pronounced for the alignment measurement. This demonstrates that the molecular beam is
considerably warmer at the sides than in the center. This is attributed to collisions with the rest
gas and to interference with mechanical apertures, \ie, skimmers. When applying an inhomogeneous
electric field, the molecules are deflected upwards. Moreover, the molecular beam is dispersed
according to the molecules' effective dipole moments, or, correspondingly, according to their
rotational states. The most polar, lowest-energy rotational states are deflected the most, and this
is reflected in the increased degrees of alignment and orientation in the deflected part of the
molecular beam. This accounts for the larger contributions of lower quantum states which show a
large deflection. In addition, a decrease of the degree of alignment and orientation in regions
where the high rotational states remain (right part of the molecular beam profile) is observed.
Helium is not present anymore in the deflected part of the molecular beam since its trajectory is
not influenced by the electric deflector. As the degree of alignment and orientation is dependent on
the position in the dispersed molecular beam, the properties of the beam change as a function of the
vertical laser probe position. This opens up the possibility for advanced multidimensional
investigations, \ie, the observation of state-selective dynamics, in the future. It can also be used
to increase the understanding of the deflection process of complex molecules, including the
investigation of nonadiabatic couplings of rotational states by the ``slowly'' changing electric
fields. Moreover, it enables the study of the nature of ensembles of molecules in a single molecular
quantum states in mixed fields, as in a previous study on impulsive alignment and mixed-field
orientation of a nearly pure ground state ensemble of OCS~\cite{Nielsen:PCCP13:18971,
   Nielsen:PRL108:193001}.

\section{Conclusions and Outlook}
\label{sec:conclusions-outlook}

In conclusion, a high-repetition-rate experimental molecular physics setup has been developed. It
allows the preparation of very strongly aligned and oriented samples of quantum-state-selected cold
molecules at a 1~kHz repetition rate. The dependence of the degree of alignment and orientation on
the position in the molecular beam has been analyzed. Both parameters are enhanced when probing the
state-selected molecules in the deflected part of the beam. A maximum degree of alignment at 1~kHz
of $\cost=0.935$ has been obtained for a 450~ps long pulse with a peak intensity of
$0.42\times10^{12}$~W/cm$^2$. The obtained degree of orientation is $\Nuptot=0.56$. This value is
lower than what was demonstrated in previous studies~\cite{Holmegaard:PRL102:023001}, and it is
limited by the combination of the applied DC electric field in the VMI spectrometer and the pulse
duration of the alignment pulse, in accordance with previous analyses~\cite{Nielsen:PRL108:193001,
   Rouzee:NJP11:105040}.

The pulse duration of our alignment pulse is continuously tunable over a wide range from $<\!50$~fs
to $>\!500$~ps; this will allow the investigation of the influence of this duration on the
(non)adiabaticity on the alignment and orientation~\cite{Nielsen:PRL108:193001}. Preliminary
measurements show that the degree of alignment is increasing when the pulse duration is shortened,
due to the stronger peak intensity available in our setup. At the same time, revivals structures
start to appear, indicating clear nonadiabatic effects in the alignment dynamics. It is likely, that
for each molecular sample a trade-off between adiabatic and non-adiabatic driving of the alignment
can be found to ensure an optimal degree of alignment and orientation.

The demonstrated strong control over molecules at high repetition rates promises the feasibility of
novel investigations of ``weak'' processes, such as chemical dynamics, using molecular-frame
photoelectron angular distributions, photoelectron holography, or X-ray or electron diffraction
imaging. For the applied, relatively short, alignment laser pulses, only moderate pulse energies, on
the order of 10~mJ are necessary. It is envisioned, that such pulses will be available at hundreds
of kHz or even MHz repetition rates in the near future. For instance, similar setups are envisioned
for the European XFEL, where burst mode lasers with similar pulse energies are under consideration
and electric state selectors can be implemented for arbitrary repetition rate molecular beams. This
opens up the possibility of performing time resolved dynamics studies using molecular frame
photoelectron angular distributions~\cite{Hansen:PRL106:073001} or photoelectron
holography~\cite{Landers:PRL87:013002, Krasniqi:PRA81:033411}. Moreover, the controlled samples
serve as ideal targets in x-ray or electron diffraction experiments~\cite{Filsinger:PCCP13:2076,
   Reckenthaeler:PRL102:213001, Hensley:PRL109:133202}, and they hold great promise toward
attosecond and high-harmonic-generation experiments of complex molecules, where the molecular
alignment and orientation can be exploited to modulate and enhance the
output~\cite{Velotta:PRL87:183901}. Moreover, the state-selection process inherently separates
structural isomers~\cite{Filsinger:PRL100:133003, Filsinger:ACIE48:6900}, which is a necessary
ingredient to investigate ultrafast dynamics of structural isomers, such as charge migration in
various conformers of glycine~\cite{Kuleff:CP338:320}.

\bigskip
\begin{acknowledgments}
   We thank the DESY-FS infrastructure groups for support, HASYLAB/Petra~III for hosting our
   laboratory, the CFEL staff for administrative and technical support and Vinod Kumarappan for the
   camera software. This work has been supported by the excellence cluster ``The Hamburg Centre for
   Ultrafast Imaging - Structure, Dynamics and Control of Matter at the Atomic Scale'' of the
   Deutsche Forschungsgemeinschaft. N.\,L.\,M.\,M.\ acknowledges financial support by the Joachim
   Herz Stiftung.
\end{acknowledgments}

%

\end{document}